\documentclass[]{pasj01}

\Received{}
\Accepted{}
 
\usepackage{lscape} 
 
\begin{document} 

\title{ALMA band 8 observations of DLA 2233+131 at $z$ = 3.150}

\author{Kazuyuki \textsc{Ogura}\altaffilmark{1}}
\altaffiltext{1}{Faculty of Education, Bunkyo University, 3337 Minamiogishima, Koshigaya, Saitama, 343-8511, Japan}
\email{ogurakz@koshigaya.bunkyo.ac.jp}

\author{Hideki \textsc{Umehata},\altaffilmark{2}}
\altaffiltext{2}{RIKEN Cluster for Pioneering Research, 2-1 Hirosawa, Wako, Saitama 351-0198, Japan}

\author{Yoshiaki \textsc{Taniguchi}\altaffilmark{3}}
\altaffiltext{3}{The Open University of Japan, 2-11 Wakaba, Mihama-ku, Chiba 261-8586, Japan}

\author{Yuichi \textsc{Matsuda}\altaffilmark{4}}
\altaffiltext{4}{National Astronomical Observatory of Japan, 2-21-1 Osawa, Mitaka, Tokyo 181-8588, Japan}

\author{Nobunari \textsc{Kashikawa}\altaffilmark{5}}
\altaffiltext{5}{Department of Astronomy, Graduate School of Science, The University of Tokyo, 7-3-1 Hongo, Bunkyo, Tokyo 113-0033, Japan}

\author{Kartik \textsc{Sheth}\altaffilmark{6}}
\altaffiltext{6}{NASA Headquarters, Washington, DC 20546-0001, USA}

\author{Katsuhiro \textsc{Murata}\altaffilmark{7}}
\altaffiltext{7}{Department of Physics, Tokyo Institute of Technology, 2-12-1 Ookayama, Meguro-ku, Tokyo 152-8551, Japan}

\author{Masaru \textsc{Kajisawa}\altaffilmark{8, 9}}
\altaffiltext{8}{Graduate School of Science and Engineering, Ehime University, Bunkyo-cho, Matsuyama, Ehime 790-8577, Japan}
\altaffiltext{9}{Research Center for Space and Cosmic Evolution, Ehime University, 2-5 Bunkyo-cho, Matsuyama, Ehime 790-8577, Japan}

\author{Masakazu A. R. \textsc{Kobayashi}\altaffilmark{10}}
\altaffiltext{10}{Faculty of Natural Sciences, National Institute of Technology, Kure College, 2-2-11, Agaminami, Kure, Hiroshima, 737-8506, Japan}

\author{Takashi \textsc{Murayama}\altaffilmark{11}}
\altaffiltext{11}{Astronomical Institute, Graduate School of Science, Tohoku University, Aramaki, Aoba, Sendai 980-8578, Japan}

\author{Tohru \textsc{Nagao}\altaffilmark{9}}



\maketitle

\begin{abstract}

We present our ALMA Band 8 observations of a damped Ly$\alpha$ absorption (DLA) system at $z$ = 3.150 observed 
in the spectrum of the quasar Q2233+131 at $z$ = 3.295. The optical counterpart of this DLA has been identified and 
it shows a double-peaked Ly$\alpha$ emission line. Since one possible origin of DLAs at high redshift is an outflowing 
gas from star-forming galaxies, DLA2233+131 provides a good laboratory to investigate the nature of high-$z$ DLAs. 
Motivated by this, we have carried out ALMA band 8 observations to study the [C II] line in this system. However, we do 
not detect any significant emission line in the observed pass bands.
 
Instead, we have serendipitously found three submm continuum sources in the observed sky area. One appears to be 
the quasar Q2233+131 itself while the other two sources are newly identified submm galaxies (SMGs), called SMG1 
and SMG2 in this paper. They are located at a separation of 4$^{\prime\prime}$.7 and 8$^{\prime\prime}$.1 from 
Q2233+131, respectively. Their 646 $\mu$m fluxes are 6.35 mJy and 6.43 mJy, respectively, being higher than that of 
Q2233+131, 3.62 mJy. Since these two SMGs are not detected in the optical images obtained with the Hubble Space 
Telescope and the Subaru Telescope, they have a very red spectral energy distribution. It is, therefore, suggested that 
they are high-redshift galaxies or very dusty galaxies at intermediate redshift although we cannot rule out the possibility 
that they are optically very faint SMG analogs at low redshift. 
Follow up observations will be necessary to explore the nature of this interesting region.

\end{abstract}


\section{Introduction} 

Quasar absorption line systems in ultraviolet (UV) spectra of high-redshift quasars have been used to investigate the nature 
of the intergalactic medium (IGM) and intervening objects such as gas-rich galaxies along lines of sight of quasars 
\citep{1986ApJS...61..249W, 2005ARA&A..43..861W}. Among such quasar absorption line systems, damped Ly$\alpha$ 
absorption (DLA) systems [$N$(HI) $\geq 2 \times 10^{20}$ cm$^{-2}$ \citep{1986ApJS...61..249W}] as well as Lyman limit 
systems (LLSs) [$10^{17}$ cm$^{-2} \leq N$(HI) $< 2 \times 10^{20}$ cm$^{-2}$ \citep{2003MNRAS.346.1103P}], and 
metal-line absorbers like Mg {\sc ii} absorbers \citep{1991A&A...243..344B, 1995qal..conf..139S, 2006ApJ...636..610R, 
2009AJ....138.1609M, 2011MNRAS.416.1215R} are useful in studying the evolution of gas-rich galaxies from high through 
intermediate to low redshifts. However, one serious problem is that their optical counterparts are rarely identified 
\citep{2005ARA&A..43..861W, 2017MNRAS.469.2959K}. 
Moreover, even when they are identified, their impact parameters are sometimes large, several tens kpc to more than 100 kpc; 
e.g., \citet{2011MNRAS.416.1215R, 2018MNRAS.479.2126F, 2019MNRAS.tmp.1435M}. 
{\citet{2014MNRAS.445..225C} examined the mass-luminosity (MZ) relation for DLAs (e.g., \cite{2013MNRAS.430.2680M}) 
to find that the stellar masses of optical counterparts of DLAs show a good agreement with those predicted by the MZ relation,
at least for the metal-rich DLAs. \citet{2017MNRAS.469.2959K} showed that the low detection rate of optical DLA counterparts 
and the distribution of impact parameters between DLAs and background quasars are explained by a simple scenario where 
the ``metallicity-luminosity relation'' of a DLA is given by $M_{\rm UV} = -5 \times [{\rm M/H}+0.3] - 20.8$ and the cross-section 
of DLA is proportional to the luminosity following a relation $\sigma_{\rm DLA} \propto L^{0.8}$. 

An open question for the nature of high-$z$ DLAs is how the counterpart makes the absorption feature.
The most probable candidates are disks of gas-rich galaxies including dwarf and low-surface brightness ones
(e.g., \cite{1997ApJ...487...73P, 1998ApJ...507..113P, 1999ApJ...514L..83J}). 
On the other hand, in these two decades, the following new ideas have been proposed; galactic outflows
(e.g., \cite{2000ApJ...532L..13T, 2004MNRAS.348..435N}), tidal tails in the galaxy interaction (e.g., \cite{2010MNRAS.406..445K}), 
and cosmic filaments (e.g., \cite{2017MNRAS.471.3686F}).

In the case of DLAs with a larger impact parameter, outflowing gas is a good candidate of the origin of absorption feature. 
For example, \citet{2000ApJ...532L..13T} proposed that an initial starburst in a galaxy with mass of $\sim$10$^{11}$ $M_{\odot}$ 
makes a large-scale shocked gaseous shell around the galaxy because of the superwind outflow. Since its typical radius reaches 
to $\sim$100 kpc, this can explain the observed large impact parameters of DLAs. The galaxy itself could become fainter during the 
course of its passive evolution making it difficult to identify optical counterparts in many cases which would then explain the observed 
low detection rate of DLA counterparts in the optical. The metal enrichment by Type II supernovae gives a typical metallicity of 
$Z \sim 0.02 Z_{\odot}$ or [M/H] $\sim -1.7$ in the shocked shell, consistent with the observed metallicity of DLAs. 
According to \citet{2001ApJ...547..146T}, the shock velocity ranges from $\sim$0 km s$^{-1}$ to a few thousands km s$^{-1}$,
depending on the energy of a superwind, the evolutionally phase of a superwind, and the viewing angle toward a part of the shocked shell.
They also found a positive correlation between the metallicity and velocity width, being consistent with the observed 
velocity-metallicity relation of high-$z$ DLAs (e.g., \cite{2006A&A...457...71L}).
Such superwind models can also explain the observed H~{\sc i} column density distribution function \citep{2004MNRAS.348..435N}.
Although it is unclear whether or not the galactic outflow is a major origin of high-$z$ DLAs, some DLAs at high redshift with
evidence for outflows have been observed so far (e.g., \cite{2004A&A...417..487C, 2013MNRAS.433.3091K, 2014ApJ...780..116K}).
For understanding the nature of high-$z$ DLAs, it is important to confirm the plausible origin of each DLA. 

Optical observations even with the largest and most sensitive telescopes are extremely challenging because these systems are extremely 
faint in the optical due to cosmological surface brightness dimming whereas millimeter/submillimeter windows offer a relatively constant 
flux due to the negative K-correction.
Therefore, instead of the optical window, we are now able to choose the millimeter/submillimeter windows using the Atacama Large Millimeter and 
Submillimeter Array (ALMA). In fact, recently, ALMA has been used to investigate the nature of high-redshift DLA counterparts by using 
dust continuum emission and radio atomic and molecular emission lines such as [C~{\sc ii}] 158 $\mu$m and CO emission lines 
\citep{2016ApJ...820L..39N, 2017Sci...355.1285N, 2018MNRAS.479.2126F, 2018ApJ...856L..23K, 2018MNRAS.474.4039M, 
2018ApJ...856L..12N, 2019ApJ...870L..19N}. 
Encouraged by these pioneering ALMA observations of high-redshift DLAs, in this study, we focus on a $z=3.150$ DLA system in the 
line of sight of Q2233+131 whose redshift is $z = 3.295$ \citep{2004A&A...417..487C}.
Because the Ly$\alpha$ emission from this optical counterpart shows a double-peaked profile, a possible origin of this DLA is thought to 
be outflowing gas. Since this DLA has been well studied by optical observations \citep{1996Natur.382..234D, 2004A&A...417..487C, 
2007A&A...468..587C, 2014ApJ...780..116K}, it provides a good laboratory to examine the scenario for the origin of the DLA, by 
combining optical and submm properties. 

The structure of this paper is as follows. In Section 2, we summarize general properties of our target, DLA2233+131. Section 3 describes
the observations and data reduction. The results of our ALMA observations are shown in Section 4. In Section 5, we discuss possible 
scenarios to explain the origin of DLA2233+131 and serendipitously detected submm galaxies (SMGs).
Finally, we present our concluding remark in Section 6. Throughout this paper, we adopt a flat cosmology with 
$\Omega_{\rm m}$ = 0.3, $\Omega_{\Lambda}$ = 0.7, and $H_0$ = 70 km s$^{-1}$ Mpc$^{-1}$.

\section{Our target: DLA2233+131} 

Our target is DLA2233+131 found toward the quasar Q2233+131 at $z=3.295$ (see Table \ref{tab:info}). 
This is the first intervening DLA whose optical counterpart was identified in both Ly$\alpha$ emission and stellar continuum 
by \citet{1996Natur.382..234D}. 
The H~{\sc i} column density of DLA2233+131 has been measured as follows; log$N$(H~{\sc i}) (cm$^{-2}$) = 20.00 
\citep{1993ApJS...84....1L}, 20.2 \citep{2002PASA...19..455C}, and 19.95 [SDSS DR5; see \citet{2014ApJ...780..116K}]. 
Since all these values are smaller than log$N$(H~{\sc i}) (cm$^{-2}$) = 20.3 required to define the DLA, strictly speaking, 
this absorber is not a DLA but a sub-DLA or a Lyman limit system (LLS); see \citet{2003MNRAS.346.1103P}. 
However, following \citet{1996Natur.382..234D}, we refer this absorber as DLA2233+131 in this paper.
Its optical image obtained with the Hubble Space Telescope is shown in the right panel of Figure \ref{fig:ALMA_HST}.

This DLA is located 2$^{\prime\prime}$.3 away from the quasar at PA = 159$^{\circ}$. 
The projected separation from the line of sight to Q2333+133 is  $\sim$ 18 kpc.
A summary of the position of DLA2233+131 measured in the literature is given in Table \ref{tab:info}.
Note that the presence of this DLA feature was 
already found by \citet{1989ApJS...69..703S} and by  \citet{1993ApJS...84....1L} and the quasar Q2233+131 was found 
in the CFHT Blue Grens Quasar Survey Plates by \citet{1985AJ.....90..987C}.
  
After the optical identification of DLA2233+131 by \citet{1996Natur.382..234D}, a number of follow up studies have been 
made to date. First, its metallicity is found as [Fe/H] = $-$1.4 \citep{1997ApJ...484..131L}.
\citet{2019arXiv190805362M} re-derived the metallicity to be [Si/H] = $-0.97 \pm 0.13$. Since Fe tends to be depleted 
by dust (e.g., \cite{2005ARA&A..43..861W, 2014ApJ...782L..29R}), we use [Si/H] as the metallicity of DLA2233+131
throughout this paper.
The Ly$\alpha$ emission-line region is further investigated by \citet{2004A&A...417..487C}
and \citet{2014ApJ...780..116K}. \citet{2004A&A...417..487C} performed integral field spectroscopic (IFS) observations of this 
DLA2233+131 to find a spatially extended Ly$\alpha$ blob with a size of $\sim$23 kpc $\times$ 38 kpc. 
However, the extended morphology of Ly$\alpha$ emission was not confirmed by higher spectral resolution
IFS observations \citep{2007A&A...468..587C}. \citet{2007A&A...468..587C} claimed that the IFS data obtained by 
\citet{2004A&A...417..487C} is likely affected by the systematic noise.
Furthermore, \citet{2014ApJ...780..116K} found not an extended but a compact Ly$\alpha$ emission-line region at DLA2233+131. 
Their analysis shows that its effective radius is 1.35 arcsec, corresponding to 10.24 kpc, and
the Sersic index is $n$ = 1.80.
\citet{2014MNRAS.445..225C} estimated the stellar mass of the counterpart of DLA2233+131 to be $\sim10^{9.8}$ $M_{\odot}$
based on their SED fitting.
The observational properties of DLA2233+131 are summarized in Table \ref{tab:counterpart}.
It is noted that another absorber at $z = 2.543$ is also identified in the UV spectrum of Q2233+131
\citep{1995AJ....110.2519S}. However, we call the absorber at $z=3.150$ as DLA2233+133 in this paper.

\begin{table} 
  \tbl{Positional information of Q2233+131 and DLA2233+131}{%
  \begin{tabular}{cccc}
      \hline
      \hline
      \multicolumn{4}{c}{Q2233+131} \\
      \hline
      $\alpha$ (J2000) & $\delta$ (J2000) & $z$ & ref. \\
      \hline
      22:36:19.19 & +13:26:20.30 & 3.295 & 1 \\
      \hline
      \hline
      \multicolumn{4}{c}{DLA 2233+131} \\
      \hline
      \multicolumn{2}{c}{Relative offset from DLA 2233+131} & & \\
      \cline{1-2} 
      $\Delta\alpha$ (arcsec) & $\Delta\delta$ (arcsec) & $z$ & ref. \\
      \hline
      +0.8 & --2.1 & 3.150\footnotemark[$*, \dag$] & 2 \\
      +1.0 & --2.1 & 3.151 & 3 \\
      +1.0 & --2.3 & 3.1501 & 4 \\
      +0.9 & --2.2 & 3.108 - 3.157\footnotemark[$\#$] & 5 \\
      \hline
  \end{tabular}}\label{tab:info}
\begin{tabnote}
{\bf References:} (1) SDSS DR5, (2) \citet{1996Natur.382..234D},
(3) \citet{1995AJ....110.2519S}; N1 in their Table 12,
(4) \citet{2002ApJ...574...51M}; N-16-1D, 
(5) \citet{2014ApJ...780..116K}. \\
\footnotemark[$*$] The value of $z_{\rm DLA}$ is taken from \citet{1993ApJS...84....1L}. \\
\footnotemark[$\dag$] $z_{\rm Ly\alpha,em}=3.1530 \pm 0.0003$ in \citet{1996Natur.382..234D}. \\
\footnotemark[$\#$] Since the narrowband filter NB502 ($\lambda_{\rm c}$ 5025 {\AA} and $\Delta\lambda$ = 60 {\AA}) is used
 in \citet{2014ApJ...780..116K}, the redshift coverage is shown here. \\
\end{tabnote}
\end{table} 

\begin{longtable}{*{4}{c}} 
\caption{Previously reported properties of the optical counterpart of DLA2233+131\label{tab:counterpart}}
\hline \hline
\multicolumn{4}{c}{Redshift ($z_{\rm DLA,em}$) of the counterpart measured from its emission line} \\
\hline
\multicolumn{2}{c}{$z_{\rm DLA,em}$} & References & Note \\
\hline 
\endfirsthead
\hline
\hline
\endhead
\hline
\endfoot
\hline
\multicolumn{4}{l}{\footnotemark[$*$] \citet{2007A&A...468..587C}.} \\
\endlastfoot
\multicolumn{2}{c}{3.1476} & \citet{2004A&A...417..487C} & \\
\hline \hline
\multicolumn{4}{c}{Impact parameter ($b$)}\\ 
\hline
\multicolumn{2}{c}{$b$} & & \\
\cline{1-2}
[arcsec] & [kpc] & References & Note \\
\hline
2.3 & 17.5 & \citet{1995AJ....110.2519S} & broad-band imaging \\
2.3 & 17.5 & \citet{1996Natur.382..234D} & long-slit spectroscopy \\
2.37 & 18.0 & \citet{2004A&A...417..487C} & IFS \\
2.41 & 18.3 & \citet{2014ApJ...780..116K} & narrow-band imaging \\
\hline \hline
\multicolumn{4}{c}{Photometric measurements} \\
\hline
\multicolumn{2}{c}{Measurements} & References & Note \\
\hline
\multicolumn{2}{c}{$V= 25.1$, $R=24.8$} & \citet{1995AJ....110.2519S} & \\
\multicolumn{2}{c}{$H= 25.34$} & \citet{2001MNRAS.326..759W} & HST/NICMOS\\
\multicolumn{2}{c}{$V_{50}= 25.75$} & \citet{2002ApJ...574...51M} & $V_{50}$ is a $V$-band filter of HST/STIS \\
\multicolumn{2}{c}{$V>24.40$, $R>24.44$, $N502=24.30$} & \citet{2014ApJ...780..116K} & N502 is a narrow-band filter of Subaru/FOCAS \\
 & & & ($\lambda_{\rm eff} = 5025$ {\AA}, $FWHM=60$ {\AA}) \\
\hline \hline
\multicolumn{4}{c}{Ly$\alpha$ flux measurements} \\
\hline
\multicolumn{2}{c}{Measurements} & References & Note \\
\hline
\multicolumn{2}{c}{$(6.4 \pm1.2) \times 10^{-17}$ [erg s$^{-1}$]} & \citet{1996Natur.382..234D} & broad-band imaging \\
\multicolumn{2}{c}{$(2.8 \pm0.3) \times 10^{-16}$ [erg s$^{-1}$]}  & \citet{2004A&A...417..487C} & IFS, overestimated\footnotemark[$*$] \\
\multicolumn{2}{c}{24.30 [AB mag]} & \citet{2014ApJ...780..116K} & narrow-band imaging \\
\end{longtable} 

\section{Observations and data reduction} 

We observed a sky area with a radius of 11$^{\prime\prime}$ centered at the position of Q2233+131 in ALMA Cycle 5 
(Proposal ID 2017.1.00345.S, PI: Y. Taniguchi). The observations were conducted on 24 May 2018 using forty-six 12-m antennas 
with baseline lengths of 15.0 m -- 313.7 m. We used the ALMA Band-8 receivers to detect redshifted [C~{\sc ii}] 158 $\mu$m and 
continuum emissions from the counterpart of DLA at $z = 3.150$ along the sightline to Q2233+131. We used 4 bands with a width 
of 1.875 GHz which are divided into 128 channels. We set the central frequency of the band to be 457.329 GHz and 459.121 GHz 
for the [C~{\sc ii}] line, while 469.329 GHz and 471.121 GHz for the continuum emission. The total on-source integration time was 
31.7 minutes.

All the data were reduced by using the Common Astronomy Software Application (CASA) ver 5.4.0 \citep{2007ASPC..376..127M}. 
To search for the emission line, we create a dirty cube per 100 km s$^{-1}$ bin. We then applied uvcontsub task for each side band 
to make the final cube with the natural weighting. As for the continuum emission, we first created a dirty map with natural weighting 
using all the 4 spectral windows. We then performed tCLEAN down to 2$\sigma$, resulting that we detected 3 continuum sources, one 
of which is Q2233+131. We detected no other objects with $>3\sigma$.
Here, the 1$\sigma$ noise level is $\sim$67 $\mu$Jy/beam at the phase center. The synthesized beam size 
is 0$^{\prime\prime}$.84 $\times$ 0$^{\prime\prime}$.61 with a position angle of 52 degree. The maximum recoverable scale is 
5$^{\prime\prime}$.15. Flux measurements were performed via Gaussian fitting with the CASA task, imfit. We extracted spectra of 
these 3 continuum sources by adopting 1$^{\prime\prime}$.0 diameter aperture.

\section{Results} 
\subsection{The ALMA Band 8 Image} 

In Figure \ref{fig:cont_map}, we show the obtained ALMA Band 8 image around the quasar Q2233+131. The field of view shown in 
Figure \ref{fig:cont_map} is approximately 22$^{\prime\prime}$ in diameter. However, note that the sky area in which the primary 
beam response exceeds 50 percent is 13$^{\prime\prime}$ in diameter (the inner dotted circle in Figure \ref{fig:cont_map}), and that 
above 30 percent is 16$^{\prime\prime}$.5 in diameter (the outer dashed circle in Figure \ref{fig:cont_map}).

As shown in this figure, we have detected three bright continuum sources. One is the quasar Q2233+131 itself that is located in the 
center of the field of view. Since the observed optical rest-frame magnitudes of Q2233+131, $B$ = 18.29 and $V$ = 18.15 
\citep{1996Natur.382..234D} are brighter by $\sim$7 magnitude than those of DLA2233+131 (see Table \ref{tab:counterpart}), we 
assume that the most submm emission comes not from DLA2233+131 but from Q2233+131. In Figure \ref{fig:ALMA_HST}, we 
compare ALMA band 8 and HST images around Q2233+131. We find no continuum source at the position of the optical counterpart 
of DLA2233+131.

The other two continuum sources are unknown ones that are considered to be SMGs although their redshifts are also unknown. 
Hereafter, we call them SMG1 and SMG2 in this paper. Their angular distances from Q2233+131 are 4$^{\prime\prime}$.7  and 
8$^{\prime\prime}$.1, respectively.
In Table \ref{tab:flux}, we give basic observational properties of the above three bright continuum sources. We here note that the 
newly found SMGs are brighter by a factor of two than the quasar Q2233+131.

\begin{figure} 
 \begin{center}
  \includegraphics[width=8cm]{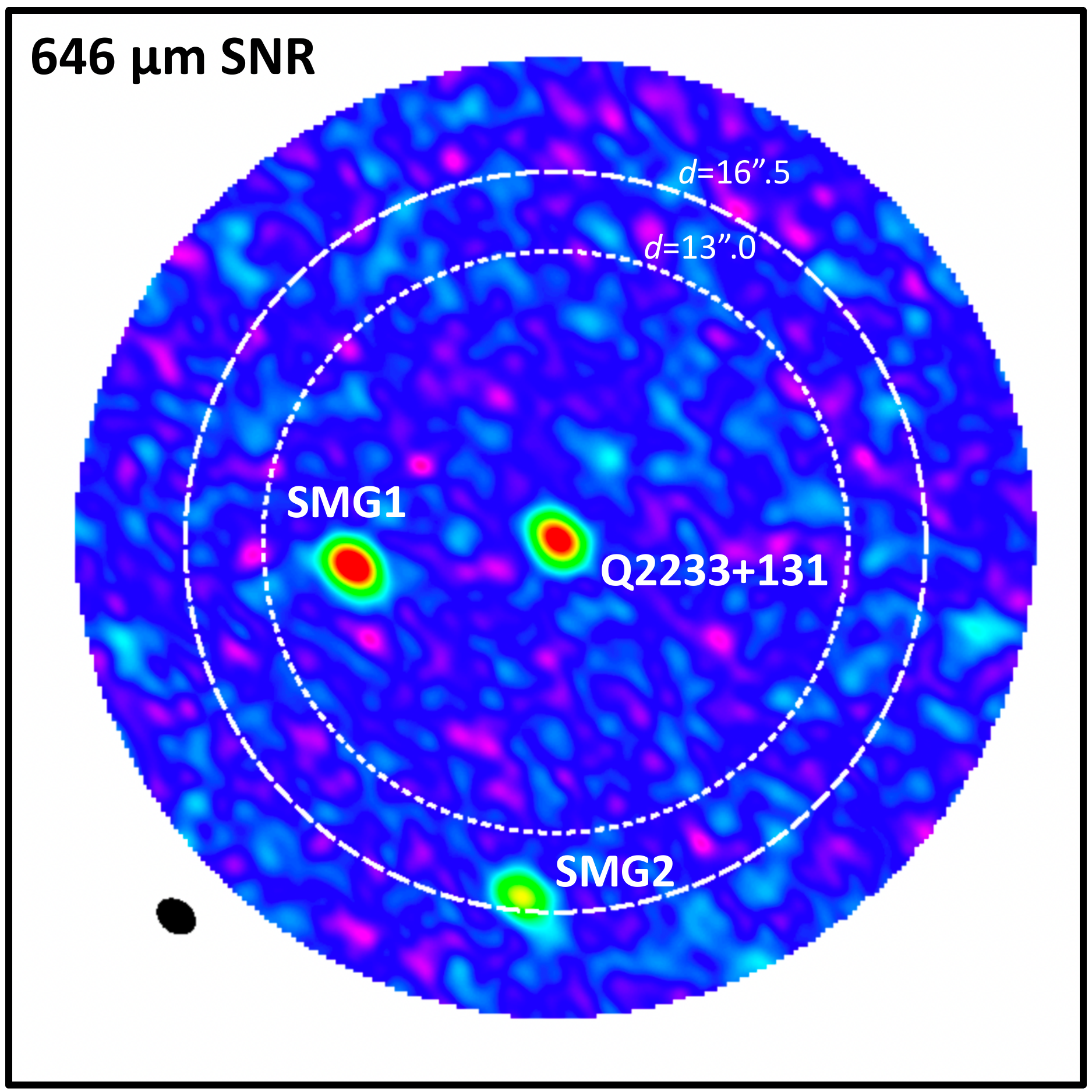} 
 \end{center}
\caption{
The ALMA rest-frame 646$\mu$m continuum map (S/N) of the DLA2233131/Q2233+131 field
(north is up and east is left).
The field-of-view of the image is $\sim$22$^{\prime\prime}$ in diameter. The dotted inner and dashed outer 
circles indicate the area where the primary beam response exceeds 50 and 30 percent, respectively. 
The black field circle in the lower left corner indicates the FWHM size of the PSF  
($0^{\prime\prime}.84 \times 0^{\prime\prime}.61$ with a position angle of 52 degree).
}\label{fig:cont_map}
\end{figure} 

\begin{figure} 
 \begin{center}
  \includegraphics[width=8cm]{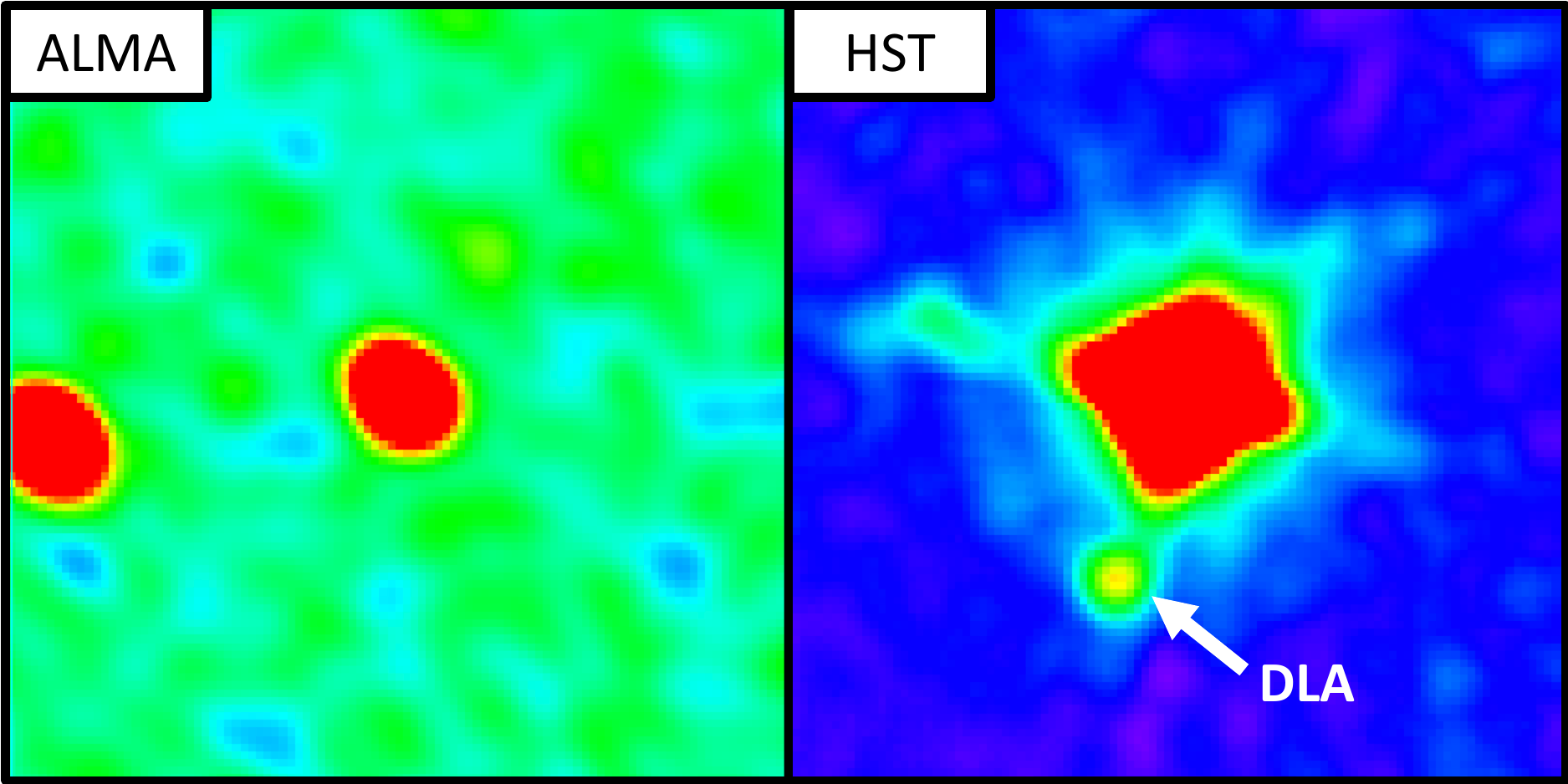} 
 \end{center}
\caption{Comparison of our ALMA image with the HST optical image around 
Q2233+131 ($10^{\prime\prime} \times 10^{\prime\prime}$).
The ALMA rest-frame 646$\mu$m continuum map (left) is compared with the 
HST WFPC2 F702W image (right) around DLA2233+131/Q2233+131. 
The ALMA source shown at the left edge in the left panel is SMG1 
(see Figure \ref{fig:cont_map}).
}\label{fig:ALMA_HST}
\end{figure} 

\begin{table} 
  \tbl{Flux measurements}{%
  \begin{tabular}{cccc}
      \hline
      Objects & R.A. & Dec. & $S_{646 \mu\rm m}$ [mJy]\footnotemark[$*$] \\ 
      \hline
      Q2233+131 & 22:36:19.19 & +13:26:20.30 & 3.62\footnotemark[$\dag$] $\pm$0.10 \\
      SMG 1 & 22:36:19.51 & +13:26:19.71 & 6.35$\pm$0.17 \\
      SMG 2 & 22:36:19.25 & +13:26:12.29 & 6.43$\pm$0.47 \\
      \hline
    \end{tabular}}\label{tab:flux}
    \begin{tabnote}
\footnotemark[$*$] Integrated flux density.  \\ 
\footnotemark[$\dag$] This value is thought to come not from DLA2233+131 but mostly from Q2233+131 (see text) .\\ 
    \end{tabnote}
\end{table} 

\subsection{The ALMA Band 8 Spectra of DLA2233+131, SMG1, and SMG2} 

In Figure 3, we show the  obtained spectrum of  DLA2233+131 together with those of  Q2233+131, SMG1, and SMG2. 
None of these objects show significant emission lines at any observed  frequency intervals. Here we use a 1$^{\prime\prime}$ 
aperture for all these sources. Since the frequency coverages of our band 8 observations are from 456.5 GHz to 459.8 GHz 
and from 468.5 GHz to 471.8 GHz, our non-detection means that [C~{\sc ii}] 158 $\mu$m emission does not appear at the 
redshift intervals of $z$ = 3.028 - 3.057 and $z$ = 3.133 - 3.163.

Now, let us estimate the upper limit of  [C~{\sc ii}] 158 $\mu$m emission line using a velocity integrated map. Here, we use the 
following three methods. Method A: Aperture photometry at the Q2233+131 position with an aperture of 0$^{\prime\prime}$.5 in 
radius, Method B: Standard deviation of 300-point random aperture photometry with an aperture of 0$^{\prime\prime}$.5 in radius, 
and Method C: RMS  of the map in units of  mJy/beam. To carry out these estimates, we first combine images with two velocity 
intervals, 200 km s$^{-1}$ and 700 km s$^{-1}$. These two intervals correspond to a typical FWHM for a star forming galaxy and 
the lowest FWHM of Ly$\alpha$ emission observed by \citet{2004A&A...417..487C}. The velocity center is set to be 458.184 GHz 
at which [C~{\sc ii}] 158 $\mu$m emission is expected to be detected at $z_{\rm DLA}$. The results are summarized in Table 
\ref{tab:upperlimit}. Although we do not know which is the best estimate, here we adopt the result of Method B with 
FWHM([C {\sc ii}]) = 700 km s$^{-1}$. 
In this case, the 3$\sigma$ velocity weighted upper limit is estimated as 330 mJy km s$^{-1}$.

\begin{figure} 
 \begin{center}
  \includegraphics[width=8.5cm]{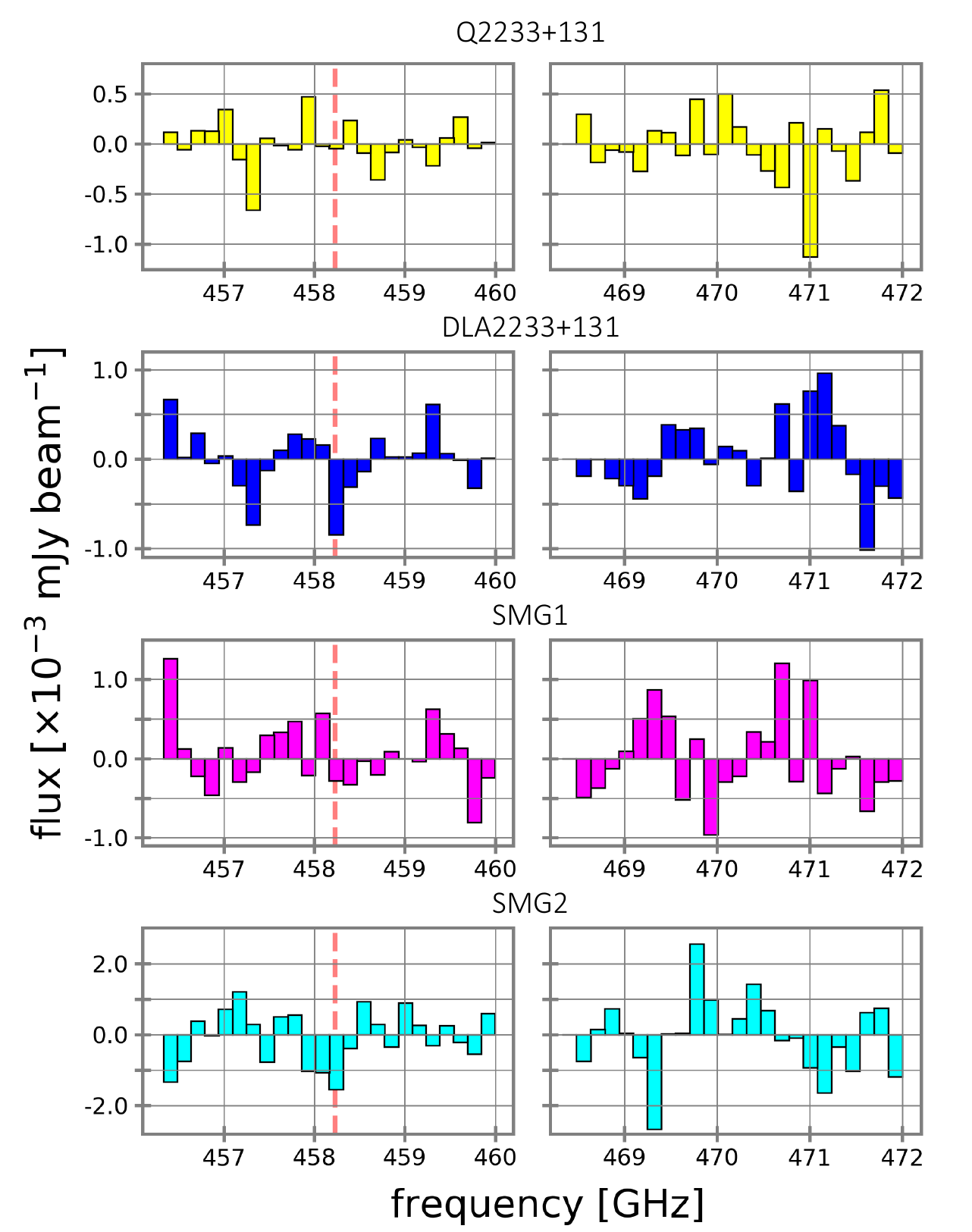} 
 \end{center}
\caption{
Spectra of  Q2233+131, DLA2233+131,  SMG1, and SMG2.
Our ALMA band 8 spectra of  Q2233+131, DLA2233+131,  SMG1 and SMG2. 
The red dashed line in each panel indicates expected observed-frame frequency 
of the [C~{\sc ii}] line at $z=3.150$.
}\label{fig:spec}
\end{figure} 

\begin{table} 
  \tbl{Various estimates of the 1$\sigma$ upper limit of  [C~{\sc ii}] 158 $\mu$m emission for DLA2233+131}{%
  \begin{tabular}{cccc}
      \hline
     FWHM([C~{\sc ii}] ) & Method A &Method B  & Method C \\ 
       & (mJy  km s$^{-1}$) &  (mJy  km s$^{-1}$)  & (mJy/beam)  \\ 
      \hline
      200 km s$^{-1}$ & 75   & 61 & 91  \\
      700 km s$^{-1}$ & 24  & 110  & 160  \\
      \hline
    \end{tabular}}\label{tab:upperlimit}
\end{table} 

\section{Discussion} 

\subsection{Why no detection in [C~{\sc ii}] from DLA2233+131 ?} 

As described in Section 4, we cannot detect [C~{\sc ii}] 158 $\mu$m emission from DLA2233+131. However, the previous ALMA 
observations detected this emission from a sample of high-$z$ DLAs at $z > 3.6$ except one target at $z =4.6$ 
\citep{2017Sci...355.1285N,  2019ApJ...870L..19N}; see Table \ref{tab:ALMA_CII}. Although their impact parameters range from 
$\sim$20 kpc to $\sim$40 kpc, the velocity coverage of [C~{\sc ii}] 158 $\mu$m emission is nearly the same as that of DLA, the 
detected [C~{\sc ii}] 158 $\mu$m emission appears to come from the object associated with the DLA in each case. 
The observed large impact parameters from a few tens to several tens kpc are significantly larger than typical sizes of such 
high-$z$ galaxies. It is thus suggested that these DLA counterparts have a large halo of dense, neutral gas around them. 

Possible origins of such large halos are considered to be either outflows from a DLA counterpart or cooling flows from primordial 
gas in cosmic webs. Our target DLA2233+131 has an evident optical counterpart observed in both Ly$\alpha$ emission line and 
UV continuum.
The presence of the double-peaked Ly$\alpha$ emission line \citep{2004A&A...417..487C} is considered as  evidence for 
outflowing gas. However, no [C~{\sc ii}] 158 $\mu$m emission is detected in our observations. Therefore, it is important to 
consider possible reasons for this non-detection.

A simple explanation is that our observations are too shallow to detect the [C~{\sc ii}] 158 $\mu$m emission line.
Indeed, previously detected [C~{\sc ii}] lines from high-$z$ DLAs are fainter than our detection limit (see Table \ref{tab:ALMA_CII}).
We estimate a probable expected flux of [C~{\sc ii}] 158 $\mu$m emission using the Ly$\alpha$ emission line flux obtained by  
\citet{1996Natur.382..234D}; $f$(Ly$\alpha$) = $(6.4 \pm 1.2) \times 10^{-17}$ erg cm$^{-2}$ s$^{-1}$.
Using the luminosity distance to DLA2233+131, $D_{\rm L}$ = 26951.9 Mpc given the cosmological parameters adopted in this 
paper, we obtain the Ly$\alpha$ luminosity, $L$(Ly$\alpha$) =  $5.5 \times 10^{42}$ erg s$^{-1}$. 
Adopting the case B recombination case, the Ly$\alpha$/H$\alpha$ intensity ratio of 8.7, we obtain  the H$\alpha$ luminosity, 
$L$(H$\alpha$) =  $6.4 \times 10^{41}$ erg s$^{-1}$. 
Based on this H$\alpha$ luminosity, we can estimate the star formation rate, $SFR \sim 5$ $M_{\odot} {\rm yr}^{-1}$
using the relation of $SFR(M_{\odot} {\rm y}^{-1}) = 7.9 \times 10^{-42} L({\rm H}\alpha)$ [erg s$^{-1}$] \citep{1998ARA&A..36..189K}. 
Using this $SFR$, we estimate the far-infrared (FIR) luminosity, 
$L$(FIR) =  $1.1 \times 10^{44}$ erg s$^{-1}$ where FIR means the wavelength interval between 1 $\mu$m and 1000 $\mu$m. 
Here, we use the relation $SFR(M_{\odot} {\rm yr}^{-1}) = 4.5 \times 10^{-44} L({\rm FIR})$ [erg s$^{-1}$] \citep{1998ARA&A..36..189K}. 
Adopting the [C~{\sc ii}] 158 $\mu$m emission to FIR luminosity ratio of $10^{-2.5}$ for $L$(FIR) $ < 10^{12} L_{\odot}$ 
\citep{2003ApJ...594..758L, 2009A&A...500L...1M}, we obtain the expected [C~{\sc ii}] 158 $\mu$m emission luminosity, 
$L$([C~{\sc ii}]) = $3.7 \times 10^{41}$ erg s$^{-1}$. 
Now, we can estimate the expected velocity integrated flux of [C~{\sc ii}] 158 $\mu$m emission as $S$([C~{\sc ii}]) 
$\delta v = L$([C~{\sc ii}]) $(1 + z) (1.04 \times 10^{-3})^{-1} \nu_{\rm rest}$([C~{\sc ii}]) $D_{\rm L}^{-2}$, where $z = 3.150$, 
$\nu_{\rm rest}$([C~{\sc ii}]) = 1900.5 GHz, and $D_{\rm L}$ = 26951.9 Mpc \citep{2005ARA&A..43..677S}. 
We then obtain $S$([C~{\sc ii}]) $\delta v$ = 302 mJy km s$^{-1}$.
This is lower than 3$\sigma$ limit of our observations (330 mJy km s$^{-1}$), as discussed in Section 4.2.

Here, it should be noted that the [C~{\sc ii}]/FIR relation of high-$z$ galaxies shows a broad scatter 
\citep{2013ARA&A..51..105C, 2017ApJ...835..110M} ranging [C~{\sc ii}]/FIR $\sim10^{-3.6} - 10^{-1.5}$. 
Based on this, the estimated [C~{\sc ii}] flux ranges $S$([C~{\sc ii}]) $\delta v$ = 24 -- 3200 mJy km s$^{-1}$. 
The non-detection of [C~{\sc ii}] emission from DLA2233+131 implies the low [C~{\sc ii}]/FIR ratio of DLA2233+131. 
\citet{2017ApJ...835..110M} showed that the main determining factor for the [C~{\sc ii}]/FIR ratio is the dust temperature;
lower values of [C~{\sc ii}]/FIR suggest higher dust temperatures.
On the other hand, we have not detected the dust continuum emission from DLA2233+131.
Moreover, \citet{2004A&A...417..487C} estimate the SFR of DLA2233+131 not only from the Ly$\alpha$ luminosity but also 
from the broad-band photometry, $SFR({\rm UV}) = 12 \pm 5$ $M_{\odot}$ yr$^{-1}$.
The $SFR({\rm Ly\alpha})$/$SFR({\rm UV})$ ratio of DLA2233+131 is comparable to those of high-$z$ Ly$\alpha$ emitters, 
suggesting that the DLA2233+131 host galaxy has only little dust content.
Therefore, there could be other possibilities for the reason why [C~{\sc ii}] emission has not detected.

\begin{longtable}{*{10}{c}} 
\caption{Summary of [C~{\sc ii}] observations of DLA with ALMA \label{tab:ALMA_CII}}
\hline \hline
$z_{\rm DLA}$ & Quasar & log$N_{\rm HI}$ & [M/H] & \multicolumn{2}{c}{$b$} & $f$(line) & FWHM & $SFR$ & ref. \\
 & & & & [$^{\prime\prime}$] &[kpc] & \tiny{[mJy km s$^{-1}$]} & [km s$^{-1}$] & $M_{\odot}$ yr$^{-1}$ & \\
\hline 
\endfirsthead
\hline
\hline
\endhead
\hline
\endfoot
\hline
\multicolumn{10}{l}{\footnotesize{\bf References:} (1) \citet{2017Sci...355.1285N}, } \\
\multicolumn{10}{l}{\footnotesize (2) \citet{2019ApJ...870L..19N}, (3) This work.} \\
\multicolumn{10}{l}{\footnotesize \footnotemark[$*$] Measured by \citet{2012ApJ...755...89R}.}  \\
\multicolumn{10}{l}{\footnotesize \footnotemark[$**$] [M/H] = [Fe/M] + 0.3 dex \citep{2012ApJ...755...89R}.}  \\
\multicolumn{10}{l}{\footnotesize \footnotemark[$\dag$] These values correspond to $\Delta V_{90}$, the velocity width} \\ 
\multicolumn{10}{l}{\footnotesize encompassing 90\% of the integrated optical depth,} \\ 
\multicolumn{10}{l}{\footnotesize (e.g., \cite{1997ApJ...487...73P}).} \\ 
\endlastfoot 
4.2584 & J0817+1351 & 21.30$\pm$0.15 & --1.15$\pm$0.15 (S)\footnotemark[$*$] & 6.2 & 42 & 13.1 & 460$\pm$50 & 110$\pm$10 & 1 \\
3.5795 & J1201+2117 & 21.35$\pm$0.15 & --0.747$\pm$0.15 (Si)\footnotemark[$*$] & 2.5 & 18 & 7.1 & 330$\pm$50 & 24$\pm$8 & 1 \\
4.3900 & J0834+2140 & 21.30$\pm$0.10 & --1.30$\pm$0.20 (S)\footnotemark[$*$] & 4.0 & 27 & 173 & 270$\pm$60\footnotemark[$\dag$]  & 7$\pm$2 & 2 \\
4.3446 & J1101+0531 & 21.00$\pm$0.20 & --1.07$\pm$0.12 (Si)\footnotemark[$*$] & 4.0 & 27 & 62 & 370$\pm$60\footnotemark[$\dag$]  & $<$7 & 2 \\ 
4.6001 & J1253+1046 & 20.30$\pm$0.15 & --1.36$\pm$0.16\footnotemark[$**$] & -- & -- & $<$42 & -- & $<$9 & 2 \\
4.2241 & PSS1443+2724 & 21.10$\pm$0.10 & --0.95$\pm$0.20\footnotemark[$**$] & 2.3 & 16 & 846 & 510$\pm$60\footnotemark[$\dag$]  & 15$\pm$4 & 2 \\
3.150 & Q2233+131 & 19.95 - 20.20 & $-0.97 \pm 0.13$ (Si) & 2.3 & 18 & $<$ 330 & -- & 22$\pm$12 & 3 \\
\hline
\end{longtable} 

A possible explanation is that the DLA counterpart is optically thick enough to activate the self-shielding effect 
\citep{2014MNRAS.445.1745W}. 
In this scenario, the [C~{\sc ii}] emission line is expected to be unobservable because the photodissociation region (PDR) 
cannot be formed. This is consistent with almost all hydrogens in systems with $N_{\rm HI} > 10^{19.5}$ are likely neutral 
\citep{1995MNRAS.276..268V}. 
Although \citet{2007A&A...468..587C} and \citet{2014ApJ...780..116K} did not confirm largely extended structure of 
DLA2233+131 reported by \citet{2004A&A...417..487C}, there appear to be some faint Ly$\alpha$ emitting regions
to the east of the galaxy associated with an idea that the DLA (see Figure 2 of \cite{2007A&A...468..587C} and Figure 
2 of \cite{2014ApJ...780..116K}).  
\citet{2004A&A...417..487C} reported the Ly$\alpha$ emission line from DLA2233+131 shows a double peak profile which 
is confirmed by the follow-up observations with a higher spectral resolution performed by \citet{2007A&A...468..587C}.
These properties suggest the existence of the outflowing gas.
If the optically thick scenario is the case, the origin of the faint Ly$\alpha$ emission in the east part of the counterpart of 
DLA2233+131 can be explained by the resonant scattering.
This scenario is similar to that proposed by \citet{2004A&A...417..487C}, where Ly$\alpha$ photons radiated by 
star-forming regions are scattered by surrounding H~{\sc i} clouds that are extended by the galactic outflow, although its
spatial extent is not so large. 

In summary, we have estimated the [C~{\sc ii}] 158 $\mu$m flux to be $S$([C~{\sc ii}]) $\delta v$ = 24 -- 3200 mJy km s$^{-1}$
based on the SFR of the counterpart. A large spread of the estimated flux is due to a broad scatter of observed [C~{\sc ii}]/FIR 
ratio. Our non-detection suggests a low [C~{\sc ii}]/FIR value and the exposure time of our observations may be too short to 
detect [C~{\sc ii}] emission from DLA2233+131.
Indeed, most of previously detected [C~{\sc ii}] emission from high-$z$ DLAs is fainter than the limiting flux of our observations
\citep{2017Sci...355.1285N, 2019ApJ...870L..19N}.
Thus it is recommended that a new observation with a longer integration time will be executed in future. 
Although there is no guideline for integration time, a couple of hours integration may be fine as a trial. 
Note that the previous successful detection of  [C~{\sc ii}] 158 $\mu$m emission from DLAs at $z \sim 4$ has been made with 
integration time of a couple of hours \citep{2017Sci...355.1285N, 2019ApJ...870L..19N}.

\subsection{What are the two SMGs ?} 

We have serendipitously found the two SMGs in the Q2233+131 field, SMG1 and SMG2 (see Fig. \ref{fig:cont_map} 
and Table \ref{tab:flux}). In this subsection, we discuss their origins.

\subsubsection{Basic properties of SMGs} 

Projected separations from SMG 1 and 2 to Q2233+131 are 4$^{\prime\prime}$.7 and 8$^{\prime\prime}$.1, respectively.
At present, we have no direct information on their redshifts. We search for the available data to date to estimate their 
redshifts. 

First, in order to search for their optical counterparts, we examine the HST/WFPC F702W and HST/NICMOS 
F160W images of the Q2233+131 field in the upper central and upper right panel of Figure \ref{fig:HSTim}, respectively. 
However, we find no optical counterpart. Note that these image were retrieved from the HST archive by 
\citet{2004A&A...417..487C} (see their Figure 7) and \citet{2001MNRAS.326..759W} (N16 in their Figure 8), respectively. 
As for HST/STIS high-resolution image for DLA2233+131, see \citet{2002ApJ...574...51M} in which DLA2233+131 is 
called as N-16-1D. However, since only a 4 arcsec by 4 arcsec area is displayed, the two SMGs are not found there. 
Next, \citet{2014ApJ...780..116K} obtained broad and narrow-band images of the Q2233+131 field to find an optical 
counterpart of DLA2233+131 using  the Faint Object Camera and Spectrograph (FOCAS; \cite{2002PASJ...54..819K})
on the Subaru Telescope. We again find no optical counterpart at the position of the SMGs in their FOCAS images 
(see the lower two panels of Figure \ref{fig:HSTim}). The 3$\sigma$ limiting AB magnitudes of FOCAS images in a 
2$^{\prime\prime}$ aperture are $B=24.40$, and $V=24.44$, respectively. 

We then checked MIR and FIR images obtained by Wide-field Infrared Survey Explorer (WISE; \cite{2010AJ....140.1868W}) 
in the framework of the ALLWISE program \citep{2014yCat.2328....0C}. Although the Q2233+131 itself is detected in all the 
WISE band images (3.4, 4.6, 12, and 22 $\mu$m), the spatial resolution of WISE 
(6$^{\prime\prime}$.1 -- 12$^{\prime\prime}$.0 depending on bands) is not high enough to resolve the quasar and the two SMGs.
Note that no image around the Q2233+131 field had been taken by the Spitzer Space Telescope.

It has been reported that most samples of previously observed SMGs are high-$z$ objects with a median redshift of $z\sim2.5$
and $\sim20$\% of SMGs are at $z>3$ (e.g., \cite{2005ApJ...622..772C, 2012MNRAS.420..957Y, 2017ApJ...840...78D}).
At low redshifts, \citet{2017arXiv170705329O} defined the $z<0.5$ analogs of SMGs based on their dust temperatures 
and total IR luminosities. All of these low-$z$ SMG analogs are detected in optical observations. However, since low-$z$ SMG 
analogs are selected by utilizing available spectral redshift, the samples may be biased toward optically bright ones. 
\citet{2012MNRAS.420..957Y} found some SMGs at $z<1$ in their deep submm imaging survey. All the SMGs at $z<1$ confirmed 
by them are brighter than $i=23.5$ mag. \citet{2017ApJ...840...78D} confirmed spectroscopic redshifts of 73 SMGs with the Keck 
Telescope. Twelve out of 73 SMGs are at $z<1$ and all of them are brighter than $R=24.2$. As for $1<z<1.5$ SMGs, 75\% (6/8) of 
them have $R<23.8$. Again, we note that these samples may be biased toward optically bright objects. 
Taking account of the non-detection of SMG1 and SMG2 in the optical with HST and the Subaru Telescope, it is more likely that 
these SMGs are located at high redshifts. 
While we found that the two SMGs are dusty and invisible at shorter wavelengths, it is hard to derive photometric redshifts for them 
due to the limited available photometric points at present.

\begin{figure*} 
 \begin{center}
  \includegraphics[width=12cm]{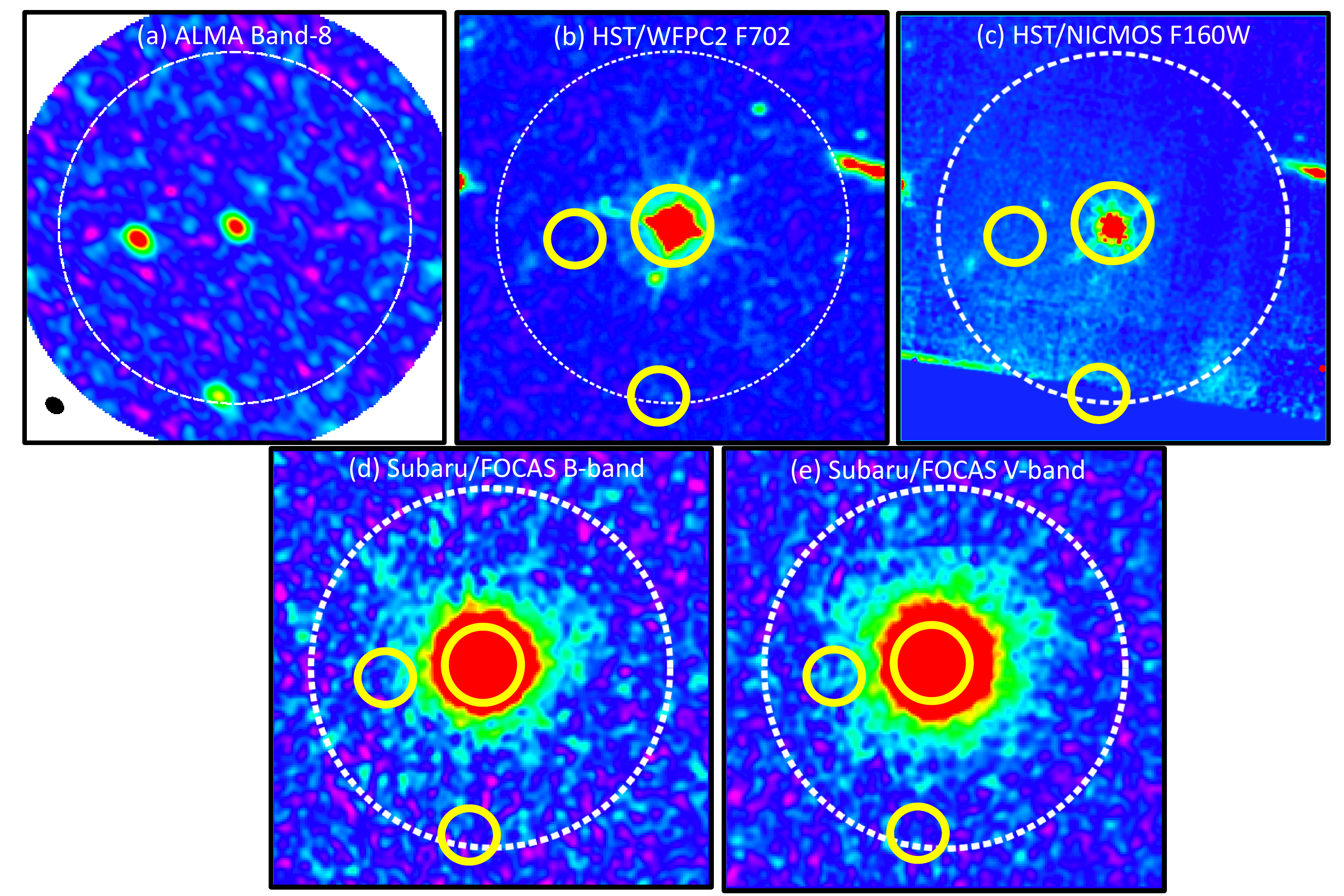} 
 \end{center}
\caption{
Comparison between ALMA band 8 image (upper left) and optical images
of the Q2233+131 field ($20^{\prime\prime} \times 20^{\prime\prime}$); 
(a) ALMA 646 $\mu$m, (b) HST WFPC2 F702W, (c) HST NICMOS F160W, (d) Subaru FOCAS B, 
and (e) Subaru FOCAS V image.
Three continuum sources, Q2233+131, SMG1, and SMG2, are shown by yellow circles.
The dashed circle in the panels (b), (c), (d), and (e), shows a sky area with a diameter of 16.5 arcsec.
}\label{fig:HSTim}
\end{figure*} 

\subsubsection{Are the two SMGs associated with Q2233+131 at $z=3.3$ ?} 

Since we cannot derive any secure redshifts of the SMGs by utilizing the current dataset, we discuss their possible association to 
Q2233+131 based on our ALMA data.
Although some submm detected DLA counterparts without optical detection have been found \citep{2019ApJ...870L..19N},
the two SMGs in this study may not be associated to DLA2233+131 at $z=3.150$ because both of them show no [C~{\sc ii}] 
emission line as shown in Figure 3.

Hence, we discuss their association to Q2233+131.

SMG1 and SMG2 are located in the sky area within a diameter of 16$^{\prime\prime}$.5, in which the primary beam response exceeds 
30 percent. In this case, the presence of the three sources, SMG1, SMG2, and Q2233+131, gives a surface number density of 
$1.4\times10^{-2}$ arcsec$^{-2}$, corresponding to $\sim$50.4 deg$^{-2}$. 
We compare this surface density with that obtained by the ALMA twenty-six arcmin$^2$ survey of GOODS-S at one millimeter (ASAGAO) 
survey \citep{2018PASJ...70..105H}. 
According to the ASAGAO survey, the surface number density of SMGs at $z\sim3$ is estimated as $\sim$900 deg$^{-2}$ for SMGs with their 
flux density of $S > 1$ mJy at $\lambda=$ 1.2 mm, giving a surface number density of $4.7  \times 10^{-4}$ arcsec$^{-2}$.
Here we estimate the flux densities at $\lambda=$ 1.2 mm of the two SMGs, SMG1 and SMG2, and the quasar Q2233+131. If we adopt 
a typical SED for SMGs at $\sim3$ \citep{2014MNRAS.438.1267S}, we obtain 1.4 mJy, 1.4 mJy, and 1.0 mJy for the three sources, respectively. 
Therefore the surface density of SMGs with their flux density of $S>1$ mJy at $\lambda=$ 1.2 mm is kept as 1.4$\times10^{-2}$ arcsec$^{-2}$. 
It is here noted that the quasar Q2233+131 is also included as an SMG. This surface number density is higher by a factor of $\sim$30 than 
that obtained with the ASAGAO project.
Note that, when we do not limit to $z\sim3$, the cumulative surface number density of ASAGAO SMGs with flux density of 
$S_{\lambda} > 1.350$ mJy (the brightest bin) at $\lambda$ = 1.2 mm is 4.2$^{+4.1}_{-2.3} \times10^{2}$ deg$^{-2}$
(see Table 4 of \cite{2018PASJ...70..105H} for further details). This is $\sim$8 times higher than the SMG surface number density of our target field. 

The obtained high surface number density with respect to that of the ASAGAO survey suggests that the two SMGs and Q2233+131 are located in 
an over density region at $z\sim3$. This also suggests that our target region comprises a so-called quasar-SMG association
(e.g., \cite{2017ApJ...844..123F, 2017ApJ...836....8T}). 
Their typical separations are $\sim10 - 100$ kpc. Such quasar-SMG systems are considered to be gas-rich merging systems 
(e.g., \cite{1999A&A...349..363G, 2013ApJ...773...44W, 2013ApJ...770...13W, 2017ApJ...844..123F}).
If SMG1 and SMG2 are located at the same redshift as that of Q2233+131, $z = 3.3$, their projected distances from Q2233+131 are 35.2 kpc and 
60.2 kpc, respectively. If this is the case, the environment around Q2233+131 provides a good laboratory for future investigations of a merging system 
at such a high redshift. We note that most of previously discovered gas-rich merging systems are identified as an unresolved single SMG by using either 
Herschel or WISE whereas they contain multiple submm sources based on ALMA observations. Here we note again that our three systems (one quasar 
with the two SMGs) are also a single source in the WISE observation. 

However, currently, there are no lines of secure evidence for their physical association. Future follow-up observations will be important to examine this 
scenario. If the SMG(s) is an optically faint object at low redshift, we may be able to measure its redshift by deep optical spectroscopic 
observations such as very sensitive long-slit or integral field spectroscopy. If the SMGs are too faint to detect even with very deep optical spectroscopy, 
ALMA is the best telescope for the redshift measurement. 
The next step will be a search for [C~{\sc ii}] 158 $\mu$m emission at the Q2233+131 redshift, $z \sim 3.3$. If still no detection, another plan is to carry 
out spectral scan observations. In this case, CO emission lines may be useful because such SMGs have been forming stars with a high SFR and thus it 
is expected that there is a lot of molecular gas in them. In this case, CO(3-2) or CO(4-3) seems a better choice, if $z \sim 2 - 3$, because this line 
can be detected at Band 3 whose observing condition is generally better than that at Band 8 in the ALMA site;  e,g.,
\citet{2018MNRAS.479.2126F} and \citet{2018ApJ...856L..12N}. 
If the SMGs are at $z<1$, the CO(2-1) line is useful to the redshift measurement (e.g., \cite{2018MNRAS.474.4039M}).

\section{Summary} 
We have conducted ALMA Band-8 observations of  DLA2233+131 at $z=3.150$ found in the UV spectrum of quasar 
Q2233+131 at $z=3.295$. 
This DLA is the first intervening DLA at high redshift whose optical counterpart was identified \citep{1996Natur.382..234D}. 
The Ly$\alpha$ emission associated with this DLA shows double peak profile whose separation corresponds to 
600 -- 750 km s$^{-1}$\citep{2004A&A...417..487C}.
High spectral resolution IFS observations by \citet{2007A&A...468..587C} and an excellent quality image obtained
by \citet{2014ApJ...780..116K} showed the optical counterpart of DLA2233+131 is a compact galaxy, although 
some faint extended Ly$\alpha$ emitting regions appear to be associated.

To investigate the origin of this DLA, we have conducted ALMA Band 8 observations to detect [C~{\sc ii}] 158 $\mu$m 
emission. However, we have not detected significant [C~{\sc ii}] emission in our ALMA observations. 
A possible reason of this non-detection is that the line is too faint to be detected in the relatively short integration 
carried out here. 
We estimate the expected [C~{\sc ii}] from DLA2233+131 based on its SFR derived by optical observations, 
$S$([C~{\sc ii}]) $\delta v$ = 24 -- 3200 mJy km s$^{-1}$. The large spread of estimated flux comes from the scatter
of observed [C~{\sc ii}]/FIR ratio. In the case that the [C~{\sc ii}]/FIR ratio is significantly low, the expected [C~{\sc ii}]
flux is fainter than the detection limit of our ALMA observations. Although such a lower [C~{\sc ii}]/FIR value suggests 
higher dust temperature \citep{2017ApJ...835..110M}, optical observations imply that DLA2233+131 may have only 
little dust content. This is consistent with the non-detection of dust continuum from DLA2233+131 in this work. 
Therefore, we consider another possible scenario in which the counterpart of DLA2233+131 is optically thick enough 
to activate the self-shielding effect. In this scenario, the PDR cannot be formed and thus [C~{\sc ii}] emission is not 
radiated. The observed double peak Ly$\alpha$ emission implies the existence of the outflowing gas. The origin of 
faint emissions around DLA2233+131 may be explained by the resonant scattering.

The previous ALMA [C~{\sc ii}] 158 $\mu$m or CO observations of low- to intermediate- to high-$z$ 
DLAs are successful \citep{2016ApJ...820L..39N, 2017Sci...355.1285N, 2018MNRAS.479.2126F, 2018ApJ...856L..23K, 
2018MNRAS.474.4039M, 2018ApJ...856L..12N, 2019ApJ...870L..19N}. 
Although we cannot detect [C~{\sc ii}] 158 $\mu$m emission from DLA2233+131, we would like to recommend that 
one carry out a systematic search for [C~{\sc ii}] 158 $\mu$m emission for a large sample of DLA from low to high redshifts. 
One problem is that it is difficult to estimate reasonable integration time because we do not have any reliable line flux 
estimators. Therefore, the first step is to carry out this survey with one hour integration for any DLAs. Such a large 
survey will tell us useful information on the origin of DLAs although there may be a variety of DLA counterparts, 
e.g., outflowing gas, disks of gas-rich galaxies from dwarf to giant ones.

Despite the non-detection of [C~{\sc ii}] 158 $\mu$m emission from DLA2233+131, we have serendipitously found the 
two SMGs in the observed field; SMG1 and SMG2. Their angular separation from Q2233+131 are 4$^{\prime\prime}$.7 
and 8$^{\prime\prime}$.1, respectively. They are not detected in the optical imaging with HST and the Subaru Telescope.
Since most of previously observed low-$z$ SMGs are also detected in optical observations, they may not be low-$z$ 
galaxies although we cannot rule out the possibility that they are optically very faint low-$z$ objects. 

Because it is hard to drive the redshifts of SMG1 and SMG2, we consider the possibility that these two SMGs and
Q2233+131 are associated with each other; i.e., SMG1 and SMG2 are at $z=3.3$ as well as Q2233+131.
If this is the case, the linear projected separations from the quasar to SMG1 and SMG2 are 35 and 60 kpc, respectively. 
These separations are similar to that of gas-rich merging systems observed as quasar-SMG pairs at high redshift
(10 -- 100 kpc). If this is the case, the Q2233+131 provides an interesting target to study gas-rich merging systems at
high redshift.
Because we cannot confine the redshifts of two SMGs, new ALMA and/or deep optical observations will be necessary 
to obtain their redshifts.


\begin{ack}
We are very grateful to the referee, Lise Chiristensen, for careful reading and many
helpful comments.
This paper uses the ALMA data ADS/JAO.ALMA\#2017.1.00345.S.
KO is grateful to Masahiro Nagasima for kindly giving many useful comments.
This work is supported from JSPS KAKENHI No. 16H02166 (PI: Y. Taniguchi).
YT would like to thank Jason X. Prochaska for his kind communication for these years.
Actually, he showed me ALMA observations made by his group prior to publication.
We would also like to thank Kotaro Kohno for useful discussion on the metallicity effect
for the detection of  [C~{\sc ii}] 158 $\mu$m emission from DLAs. 
\end{ack}



\end{document}